\documentclass[lettersize,journal]{IEEEtran}

\usepackage{cite}
\usepackage{amsmath,amssymb,amsfonts}
\usepackage{algorithmic}
\usepackage{graphicx}
\usepackage{textcomp}
\usepackage{xcolor}
\usepackage{subfigure}
\usepackage{caption}
\usepackage{ifpdf}
\usepackage{float}
\usepackage{booktabs}

\ifCLASSINFOpdf
\else
\fi
\usepackage{amsmath}
\usepackage{algorithmic}

\usepackage{array}
\begin{document}
%
\title{Performance Analysis of Hybrid RF-Reconfigurable Intelligent Surfaces Assisted FSO Communication} 
%
%
%

\author{Haibo Wang,~\IEEEmembership{Member,~IEEE,}
        Zaichen Zhang,~\IEEEmembership{Senior Member,~IEEE,} Bingcheng Zhu,~\IEEEmembership{Member,~IEEE,} and~Yidi Zhang
\thanks{Haibo Wang, Zaichen Zhang, Bingcheng Zhu, and Yidi Zhang are with National Mobile Communications Research Laboratory, Southeast University, Nanjing 210096, China. Zaichen Zhang is the corresponding author.
}

}

%
%

\markboth{Journal of \LaTeX\ Class Files,~Vol.~xx, No.~xx, xx~xxxx}%
{Shell \MakeLowercase{\textit{et al.}}: Bare Demo of IEEEtran.cls for IEEE Journals}
%



\maketitle

\begin{abstract}
Optical reconfigurable intelligent surface (ORIS) is an emerging technology that can achieve reconfigurable optical propagation environments by precisely adjusting signal's reflection and shape through a large number of passive reflecting elements. In this paper, we investigate the performance of an ORIS-assisted dual-hop hybrid radio frequency (RF) and free space optics (FSO) communication system. By jointly considering the physical models of ORIS, RF channel, atmospheric turbulence, and pointing error, the closed-form solutions of the system's precise outage probability, asymptotic outage probability and BER have been derived. It is shown through numerical results that the derivation results are accurate and the RF-FSO links with ORISs show a slightly worse performance than the traditional RF-FSO links. Based on theoretical analysis and simulation results, the system design and effect of each parameter have been discussed.
\end{abstract}


%
\IEEEpeerreviewmaketitle

\section{Introduction}\label{introduction}

\IEEEPARstart {F}{ree} space optics (FSO) systems can provide higher bandwidth, capability and security compared to traditional radio frequency (RF) systems. However, FSO systems have strong limitations, which are restricted to obstacles, weather conditions, atmospheric turbulence and pointing errors. To address this problem, hybrid FSO-RF and RF-FSO systems have been proposed, which can improve the performance in different scenarios by mixing the FSO and RF links. The difference between FSO-RF and RF-FSO systems lies in the positions of RF links and FSO links.Among them, the advantage of the RF-FSO system is to ensure that the front-end signal is not interrupted and interfered in the environment, while providing high-bandwidth and energy-efficient communication to the user.\par
Optical reconfigurable intelligent surface (ORIS), as a new type of programmable communication device, can achieve reconfigurable optical propagation environments by precisely adjusting signal's reflection and shape through a large number of passive reflecting elements, which promise to improve the performance of hybrid RF-FSO communication. In existing FSO and visible light communication (VLC), ORIS has been applied to improve the system performance. In \cite{9443170}, the performance of the ORIS-assisted FSO communication system is derived and analyzed. The single-mirror-type ORIS can change the optical path and perform beam deflection in real time. However, compared to the FSO system with a direct path, ORIS will introduce a certain amount of channel fading. In \cite{9466323}, multiple ORISs in different positions are used in the FSO system for optical diversity transmission. Analysis shows that multiple ORIS can improve the performance of the FSO system and reduce the system's bit error rate (BER) leveling in scenes containing obstacles. In the RF-FSO hybrid wireless communication system, the FSO link was originally only used as a point-to-point fixed high-speed transmission link between the relay and the user \cite{8048557,9438638}. After adding ORIS, since ORIS can deflect, split and distribute the beam, the degree of freedom of the FSO link has been further improved. Real-time beam control and space division multiplexing can be realized with multiple ORISs. However, in the existing research, there is no analysis of the ORIS-assisted RF-FSO hybrid wireless communication system.\par
This paper designs an ORIS-assisted RF-FSO hybrid wireless communication system, and performs mathematical modeling and performance analysis on it. Different from the previous work, when analyzing the performance of the RF-FSO hybrid wireless communication system, the physical models of RF channel, atmospheric turbulence, pointing error and ORIS are jointly considered in this paper. Among them, pointing error includes both the jitter of the transmitter and the beam shift caused by the jitter of the ORIS surface. The closed-form solutions of the system's precise outage probability, asymptotic outage probability and BER have been derived. The simulation results show that the deduced results are accurate. Based on theoretical analysis and simulation results, the parameter settings of the system have been discussed.\par
The rest of this paper are organized as follows. Section \ref{model} introduces our system model and derive the probability density function (PDF) of the channel fading in RF and FSO links. In Section \ref{per}, we derive the expression for system's precise outage probability, asymptotic outage probability and BER. Section \ref{num} shows some numerical results and makes some discussions on system parameters. Section \ref{conclusion} draws conclusion.

\section{System Model}\label{model}
\begin{figure}[htbp]
\centering
\includegraphics[width=0.49\textwidth]{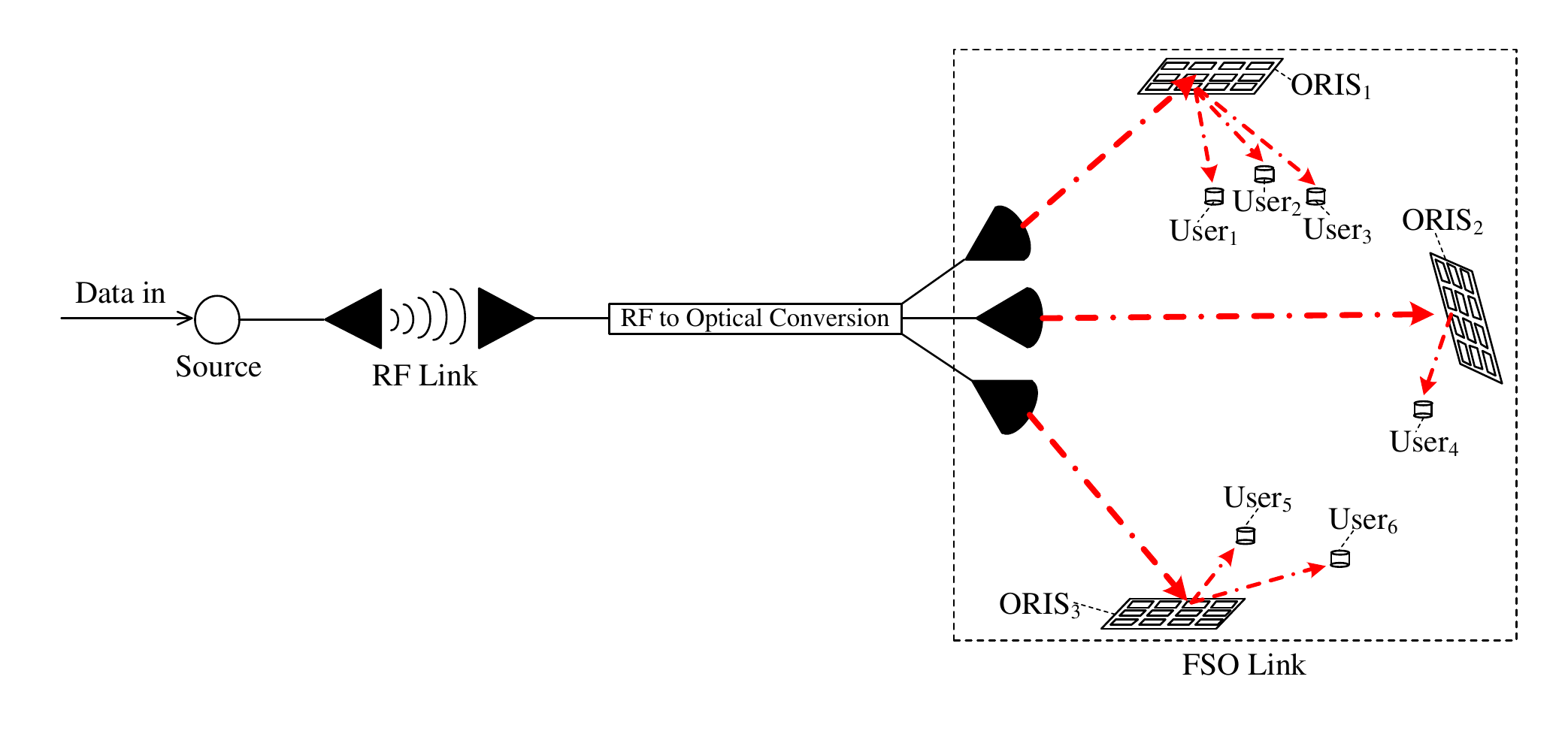}
\caption{Application scenarios of the multi-branch ORISs-assisted optical wireless communication system.}
\label{fig.1}
\end{figure}

In this work, we consider an ORIS assisted RF-optical hybrid communication system, which is shown in Fig. \ref{fig.1}. The base station transmits RF signals to the relay. The relay decodes the radio frequency signal, and then generates multiple optical signals with different information according to the address code and forwards them to multiple ORISs. ORIS performs deflection, splitting and power distribution to the beam. Since the number of light sources and ORIS is fixed, it is assumed that each ORIS is used to serve users in a sub-area. When adding or reducing users in a sub-area, the ORIS needs to be re-splitting to dynamically adjust the signal. Therefore, in this system, the number of users is not fixed. The system needs to be dynamically controlled by adjusting the ORIS and power distribution coefficient, which is discussed in Section 4. Since the RF link from the base station to the relay and the optical link channel from the relay to the user are independent of each other, we will first analyze the channels of the two links separately.
\subsection{RF transmission link}
The received signal of the relay can be written as
\begin{equation} \label{2-1}
\begin{split}
y_r=\sqrt{P_t}h_{b,r}s_{b,r}+n_r
\end{split}
\end{equation}
where $P_t$ is the transmit power of the base station, $h_{b,r}$ is the channel fading from base station to relay, $s_{b,r}$ is the transmitted symbol with $E\left \{ \left | s_{b,r} \right | ^{2}\right \}=1$, $E\left \{ \cdot  \right \}$ is the mathematical expectation, and $n_r$ is an additive white Gaussian noise term with the variance of $\sigma_{n_r}^2$.\par
Since the base station transmits a high-rate RF beam to the relay in this system, the channel contains a fixed direct component. We assume that the channel fading from the base station to the relay conforms to the Rice distribution. As $P_t=1$, the probability density function (PDF) of the envelope of of the relay's received signal $\nu$ can be written as
\begin{equation} \label{2-2}
\begin{split}
f_\nu(\nu)=\frac{\nu}{\sigma_m ^2}exp\left ( -\frac{\nu^2+A^2}{2\sigma_m ^{2}} \right )\cdot I_0\left ( \frac{\nu A}{\sigma_m ^{2}} \right )
\end{split}
\end{equation}
where $A$ is the normalized peak value of the amplitude of the main signal, $\sigma_m ^{2}$ is the normalized power of the multipath component, $I_0$ is the modified $0-th$ order Bessel function of the first kind. In additon, $K$ is the Rice factor and $K=\frac{A^2}{2\sigma_m ^{2}}$, which indicates the proportion of deterministic component.
\subsection{Optical transmission link}
Since the relay needs to extract user information from the signal, and then perform space division multiplexing according to the user's spatial location in this system, the method of decoding and forwarding is adopted. The relay receives and decodes the RF signal. Subsequently, the relay generates multiple optical signals containing different users' information and sends them to multiple ORISs. It is assumed that there are M ORISs and Q users. The signal received by the user $k$ that is served by ORIS m, can be written as
\begin{equation} \label{2-6}
\begin{split}
y_{u_k}=\alpha_m\mu_k h_{r,k}s_{r,m}+n_k
\end{split}
\end{equation}
where $\alpha_m$ is the attenuation coefficient introduced by the ORIS m, $\mu_k$ is the power allocation coefficient allocated to user $k$ ($\mu_k=\mu_{r,m}\mu_{m,k}$, where $\mu_{r,m}$ is the power allocation coefficient allocated by the relay to the ORIS m, and $\mu_{m,k}$ is the power allocation coefficient allocated by the ORIS m to the user $k$), $h_{r,k}$ is the channel fading from the relay to user $k$, $s_{r,k}$ is the signal transmitted by the relay to ORIS k, $n_k$ is the zero-mean Gaussian white noise from the user $k$'s receiver with a variance of $\sigma^2_{n_k}$. Since ORIS will change the phase of the optical signal, we utilize intensity modulation direct detection (IM/DD) with on-off keying (OOK) modulation in this system and $s_{r,k}=0$ or $2P_o$, where $P_o$ is the transmitted optical power at the relay and $P_o=\delta P_t$, $\delta$ is the optical power conversion coefficient at the relay.\par
Next, we will analyze the channel fading in the optical link, which arises due to path loss, pointing error and atmospheric turbulence in this system. The channel fading $h_{r,k}$ can be expressed as
\begin{equation} \label{2-7}
\begin{split}
h_{r,k}=h_{l_{r,k}}h_{p_{r,k}}h_{a_{r,k}}
\end{split}
\end{equation}
where $h_{l_{r,k}}$ is the channel fading caused by path loss, which is deterministic,$ h_{p_{r,k}}$, $h_{a_{r,k}}$ are channel fading caused by pointing error and atmospheric turbulence, relatively, which are random variables with distributions discussed below.\par
\subsubsection{Pointing Error}
In FSO system, pointing error refers to the deviation of the beam on the receiver plane caused by the jitter of the transmitter. For communication systems with ORIS, pointing error also includes the beam offset caused by the jitter of the ORIS surface. According to the physical model of ORIS in \cite{9443170} and \cite{9627820}, the PDF of the jitter angle at the receiving plane $\theta_u$ can be written as
\begin{equation} \label{2-8}
\begin{split}
f_{\theta_u}(\theta_u)=\frac{\theta_u}{\left(1+\frac{l_{r,o}}{l_{o,u}}\right)^2\sigma _{\theta}^{2}+  4\sigma _{\beta}^{2} }e^{-\frac{\theta_u^{2}}{2\left(1+\frac{l_{r,o}}{l_{o,u}}\right)^2\sigma _{\theta}^{2}+  8\sigma _{\beta}^{2}}}
\end{split}
\end{equation}
where $l_{r,o}$ is the link distance from the relay to ORIS, $l_{o,u}$ is the link distance from ORIS to the user, $\theta$ is the beam's jitter angle at the transmitter, $\beta$ is the jitter angle of ORIS, $\sigma _{\theta}^{2}$ is the variance of $\theta$, and $\sigma _{\beta}^{2}$ is the variance of $\beta$. As $\theta_u$ is the angle corresponding to the light beam offset in the user's receiving plane $R$, the instantaneous displacement from the receiver center to receiving light spot $R$ can be presented as
\begin{equation} \label{2-9}
\begin{split}
R=tan\theta_ul_{o,u}\approx\theta_ul_{o,u}
\end{split}
\end{equation}
From \cite{4267802}, the channel fading caused by pointing error $h_p$ can be approximated as
\begin{equation} \label{2-10}
\begin{split}
h_{p} \approx A_0\exp\left(-\frac{2R^2}{\omega_{zeq}^2}\right)
\end{split}
\end{equation}
where $A_0$ is the fraction of the receiver's collected power at $R=0$ and $\omega_{zeq}$ is the equivalent beam width. We have $A_0=[{\rm erf}(z)]^2$ and $\omega_{zeq}^2=\omega_z^2\frac{\sqrt{\pi}{\rm erf}(z)}{2zexp(-z^2)}$, where $z=\sqrt{\frac{\pi}{2}}\frac{a}{\omega_z}$ is the ratio between aperture radius and the beam width, ${\rm erf}(x)=\frac{2}{\sqrt{\pi}}\int_0^xe^{-t^2}dt$ is the error function, $\omega_z$ describes the increase of the beam radius with the propagation distance from the relay and can be approximated by $\omega_z=\phi(l_{r,o}+l_{o,u})$, $\phi$ is the divergence angle of the beam. The approximation in \eqref{2-10} is very accurate if $\frac{\omega_z}{a}>6$, where $a$ is the receiver's aperture radius \cite{4267802}.\par
Then substituting \eqref{2-9} and \eqref{2-10} into \eqref{2-8}, the PDF of $h_p$ can be derived as
\begin{equation} \label{2-11}
\begin{split}
f_{h_{p}}(h_{p})&=\frac{\omega^2_{zeq}}{4A_0\sigma _{\theta}^{2}\left ( l_{r,o}+l_{o,u} \right )^2+16A_0\sigma _{\beta}^{2}l_{r,o}^2}
\\ &\times \left ( \frac{h_p}{A_0} \right )^{\frac{\omega^2_{zeq}}{4\sigma _{\theta}^{2}\left ( l_{r,o}+l_{o,u} \right )^2+16\sigma _{\beta}^{2}l_{r,o}^2}-1},\quad 0<h_{p}<A_0,
\end{split}
\end{equation}
\subsubsection{Atmospheric Turbulence}
In this system, the relay and ORIS are located relatively close to users thus it is assumed that the atmospheric turbulence in the optical link is weak. The statistical model of FSO channel fading caused by atmospheric turbulence has been studied in depth \cite{4432329,4267802,8894089}. For weak turbulence, the intensity fluctuation is modeled as a log-normal distribution, which is experimentally verified \cite{4432329,8894089}. From \cite{4267802}, the PDF of the channel fading caused by atmospheric turbulence $h_a$ can be expressed as
\begin{equation} \label{2-12}
\begin{split}
f_{h_a}(h_a)=\frac{1}{2h_a\sqrt{2\pi\sigma _{X}^{2}}}e^{\frac{\left ( lnh_a+2\sigma _{X}^{2} \right )^{2}}{8\sigma _{X}^{2}}}
\end{split}
\end{equation}
where $\sigma _{X}^{2}$ is the log-amplitude variance, which is given by 
\begin{equation} \label{2-13}
\begin{split}
\sigma _{X}^{2}=0.30545\kappa^{\frac{7}{6}}C^2_n(L)(l_{r,o}+l_{o,u})^{\frac{11}{6}}\approx\frac{\sigma_{R}^2}{4}
\end{split}
\end{equation}
where $C^2_n(L)$ is the index of refraction structure parameter at altitude $L$, which can be assumed to be constant along the propagation path, $\kappa=\frac{2\pi}{\lambda} $ is the optical wavenumber, $\lambda$ is the optical wavelength, and $\sigma_{R}^2$ is the Rytov variance defined as
\begin{equation} \label{2-14}
\begin{split}
\sigma _{R}^{2}=1.23C^2_n\kappa^{\frac{7}{6}}(l_{r,o}+l_{o,u})^{\frac{11}{6}}
\end{split}
\end{equation}
\subsubsection{Optical Channel Fading}
The PDF of optical channel fading from the relay to the user $k$ $h_{r,k}$ can be expressed as
\begin{equation} \label{2-15}
\begin{split}
f_{h_{r,k}}(h_{r,k})=\int f_{h_{r,k}\mid h_{a_{r,k}}}\left ( h_{r,k}\mid h_{a_{r,k}} \right )f_{h_{a_{r,k}}}\left ( h_{a_{r,k}} \right )dh_{a_{r,k}}
\end{split}
\end{equation}
where $f_{h_{r,k}\mid h_{a_{r,k}}}\left ( h_{r,k}\mid h_{a_{r,k}} \right )$ is the conditional PDF given the turbulence state $h_{a_{r,k}}$, which can be derived from \eqref{2-7} and \eqref{2-11} as
\begin{equation} \label{2-16}
\begin{split}
f_{h_{r,k}\mid h_{a_{r,k}}}\left ( h_{r,k}\mid h_{a_{r,k}} \right )&=\frac{1}{h_{a_{r,k}}h_{l_{r,k}}}f_{h_{p_{r,k}}}\left( \frac{h_{r,k}}{h_{a_{r,k}}h_{l_{r,k}}}\right)\\&=\frac{\rho}{A_0h_{a_{r,k}}h_{l_{r,k}}}\left ( \frac{h_{r,k}}{A_0h_{a_{r,k}}h_{l_{r,k}}} \right )^{\rho-1},
\end{split}
\end{equation}
where $\rho=\frac{\omega^2_{zeq}}{4\sigma _{\theta}^{2}\left ( l_{r,o}+l_{o,u} \right )^2+16\sigma _{\beta}^{2}l_{r,o}^2}$.\par
Substituting \eqref{2-16} and \eqref{2-12} into \eqref{2-15}, we can obtain the closed form of the PDF of $h_{r,k}$
\begin{equation} \label{2-17}
\begin{split}
f_{h_{r,k}}(h_{r,k})&=\frac{\rho}{\left (A_0h_{l_{r,k}}  \right )^{\rho }}h_{r,k}^{\rho -1}\int _{\frac{h_{r,k}}{A_0h_{l_{r,k}}}}^{\infty }h_{a_{r,k}}^{-\rho }f_{h_{a_{r,k}}}\left ( h_{a_{r,k}} \right )dh_{a_{r,k}}\\&=\frac{\rho}{\left (A_0h_{l_{r,k}}  \right )^{\rho }}h_{r,k}^{\rho -1}\int _{\frac{h_{r,k}}{A_0h_{l_{r,k}}}}^{\infty }h_{a_{r,k}}^{-\rho }\frac{1}{2h_{a_{r,k}}\sqrt{2\pi\sigma^2_X}}\\ &\times exp\left ( \frac{\left ( lnh_{a_{r,k}}+2\sigma^2_X \right )^{2}}{8\sigma^2_X} \right )dh_{a_{r,k}} \\&=\frac{\rho}{2\left (A_0h_{l_{r,k}}  \right )^{\rho }}h_{r,k}^{\rho -1}\\&\times erfc\left ( \frac{ln\frac{h_{r,k}}{A_0h_{l_{r,k}}}+2\sigma _{X}^{2}+4\rho \sigma_{X}^{2}}{2\sqrt{2}\sigma _{X}} \right )e^{2\sigma _{X}^{2}\rho \left ( 1+\rho  \right )}.
\end{split}
\end{equation}
Then the CDF of $h_{r,k}$ can be derived as
\begin{equation} \label{2-18}
\begin{split}
F_{h_{r,k}}(h_{r,k})&=\int _{-\infty }^{x}f_{h_{r,k}}(h_{r,k})dh_{r,k}\\&=\frac{1}{2}exp\left(\rho ln\frac{h_{r,k}}{A_0h_{l_{r,k}}}+2\rho\sigma _{X}^{2}+2\rho^2\sigma_{X}^{2}\right)\\&\times erfc\left ( \frac{ln\frac{h_{r,k}}{A_0h_{l_{r,k}}}+2\sigma _{X}^{2}+4\rho \sigma_{X}^{2}}{\sqrt{8}\sigma _{X}} \right )\\&+\frac{1}{2}erfc\left ( \frac{ln\frac{A_0h_{l_{r,k}}}{h_{r,k}}-2\sigma _{X}^{2}}{\sqrt{8}\sigma _{X}} \right ),
\end{split}
\end{equation}

\section{Performance Analysis}\label{per}
\subsection{Outage Probability}
Since this system uses the decoding and forwarding method, the outage probability of the user $k$ $P_{out}(\gamma)$ can be expressed as
\begin{equation} \label{2-111}
\begin{split}
P_{out}(\gamma_{th})&=1-(1-F_{\gamma_{b,r}}(\gamma_{th}))(1-F_{\gamma_{r,k}}(\gamma_{th}))\\
&=F_{\gamma_{b,r}}(\gamma_{th})+F_{\gamma_{r,k}}(\gamma_{th})-F_{\gamma_{b,r}}(\gamma_{th})F_{\gamma_{r,k}}(\gamma_{th})
\end{split}
\end{equation}
which means that the communication system will not be interrupted only when the RF and optical channels are both uninterrupted, where $F_{\gamma_{b,r}}(\cdot)$ is the cumulative distribution function (CDF) of the RF channel's signal-to-noise ratio (SNR) from the base station to the relay, and $F_{\gamma_{r,k}}(\cdot)$ is the CDF of the optical channel's SNR from the relay to the user $k$, $\gamma_{th}$ is the outage threshold of SNR.\par
\subsubsection{RF Link}
From \eqref{2-1}, the SNR of the relay's received signal can be expressed as
 \begin{equation} \label{2-112}
\begin{split}
\gamma_{b,r}=\frac{P_t\nu^2}{\sigma_{n_r}^{2}}
\end{split}
\end{equation}
Then the PDF of $\gamma_{b,r}$ can be derived from \eqref{2-2} as
\begin{equation} \label{2-3}
\begin{split}
f_{\gamma_{b,r}}(\gamma_{b,r})&=\frac{\sigma^2_{n_r}}{2P_t\sigma^2_m}exp\left ( -\frac{\sigma_{n_r} ^{2}\frac{\gamma_{b,r}}{P_t}+A^{2}}{2\sigma_m ^{2}} \right )\\&\times I_0\left ( \frac{\sigma_{n_r}A\sqrt{\gamma_{b,r}}}{\sigma_m^2\sqrt{P_t}} \right )
\end{split}
\end{equation}
Then the cumulative distribution function (CDF) of $\gamma_{b,r}$ can be derived as
\begin{equation} \label{2-4}
\begin{split}
F_{\gamma_{b,r}}(x)&=1-exp\left ( -\frac{\sigma_{n_r} ^{2}x+A^{2}P_t}{2\sigma_m ^{2}P_t} \right )\\&\times \sum_{k=0}^{\infty }\left ( \frac{A}{\sigma _{n_r}}\sqrt{\frac{P_t}{x}} \right )^{k}I_{k}\left ( \frac{A\sigma_{n_r}}{\sigma _{m}^2}\sqrt{\frac{x}{P_t}} \right )\\&=1-Q_1\left(\frac{A}{\sigma_m},\frac{\sigma_{n_r}}{\sigma_m}\sqrt{\frac{x}{P_t}}\right),
\end{split}
\end{equation}
where $Q_1(\cdot)$ is the Marcum Q-function.
\subsubsection{Optical Link}
According to \eqref{2-6}, the instantaneous SNR in the optical channel from the relay to the user $k$ $\gamma_{r,k}$ can be defined as
\begin{equation} \label{2-19}
\begin{split}
\gamma_{r,k}=\frac{2\mu _{k}^2\alpha _{m}^{2}h_{r,k}^{2}\delta^2P_{t}^{2}}{\sigma _{n_k}^{2}}.
\end{split}
\end{equation}
Substituting \eqref{2-19} into \eqref{2-18}, the CDF of $\gamma_{r,k}$ can be derived as
\begin{equation} \label{2-20}
\begin{split}
&F_{\gamma_{r,k}}(\gamma_{r,k})=\frac{1}{2}exp\left(\rho \eta+2\rho\sigma _{X}^{2}+2\rho^2\sigma_{X}^{2}\right)\\&\times erfc\left ( \frac{\eta+2\sigma _{X}^{2}+4\rho \sigma_{X}^{2}}{\sqrt{8}\sigma _{X}} \right )-\frac{1}{2}erfc\left (\frac{\eta+2\sigma _{X}^{2}}{\sqrt{8}\sigma _{X}} \right ),
\end{split}
\end{equation}
where $\eta=\frac{1}{2}ln\frac{\gamma_{r,k}}{2}+ln\frac{\sigma_{n_k}}{A_0P_t\delta\alpha_m\mu_kh_{l_{r,k}}}$.\par
Substituting \eqref{2-4} and \eqref{2-20} into \eqref{2-19}, we can obtain the outage probability of the user $k$ as
\begin{equation} \label{2-21}
\begin{split}
&P_{out}(\gamma_{th})=1+\frac{1}{2}exp\left(\rho \eta^{'}+2\rho\sigma _{X}^{2}+2\rho^2\sigma_{X}^{2}\right)\\ &\times erfc\left ( \frac{\eta^{'}+2\sigma _{X}^{2}+4\rho \sigma_{X}^{2}}{\sqrt{8}\sigma _{X}} \right )Q_1\left(\frac{A}{\sigma_m},\sqrt{\gamma_{th}}\right)\\&-\frac{1}{2}erfc\left (\frac{\eta^{'}+2\sigma _{X}^{2}}{\sqrt{8}\sigma _{X}} \right )Q_1\left(\frac{A}{\sigma_m},\sqrt{\gamma_{th}}\right)-Q_1\left(\frac{A}{\sigma_m},\sqrt{\gamma_{th}}\right),
\end{split}
\end{equation}
where $\eta^{'}=\frac{1}{2}ln\frac{\gamma_{th}}{2}+ln\frac{\sigma_{n_k}}{A_0P_t\delta\alpha_m\mu_kh_{l_{r,k}}}$.
\subsection{Asymptotic Analysis}
\subsubsection{Asymptotic Outage Probability and BER}
In this work, we derive the asymptotic performance expressions based on the analytical technique in \cite{1221802}. First, we expand the \eqref{2-21} and keep the low-order terms of $\frac{1}{\overline{\gamma}}$, where $\overline{\gamma}$ represents the average SNR, the aymptotic outage probability can be estimated as
\begin{equation} \label{3-5}
\begin{aligned}
P_{out}^{\infty}(\gamma_{th})&\approx \frac{\gamma_{th}^{\frac{\rho }{2}}}{2^{\frac{\rho }{2}+1}}\left ( \frac{\sigma_{n_k}}{A_0P_t\delta\alpha_m\mu_kh_{l_{r,k}}} \right )^{\rho }e^{2\rho \sigma _{X}^{2}+2\rho ^{2}\sigma _{X}^{2}}\\&+\frac{A^4\sigma_{n_r}^2+4\sigma _{m}^{4}\sigma_{n_r}^2-2\sigma _{m}^{2}\sigma_{n_r}^2A^{2}}{8\sigma _{m}^{6}P_t}\gamma _{th}
\end{aligned}
\end{equation}
Then according to \cite{1221802}, the aymptotic PDF of system's SNR $\gamma$ can be derived as
\begin{equation} \label{3-6}
\begin{aligned}
f_{\gamma}(\gamma)&\approx \frac{\rho\gamma^{\frac{\rho }{2}-1} }{2^{\frac{\rho }{2}+2}} \left ( \frac{\sigma_{n_k}}{A_0P_t\delta\alpha_m\mu_kh_{l_{r,k}}} \right )^{\rho }e^{2\rho \sigma _{X}^{2}+2\rho ^{2}\sigma _{X}^{2}}\\&+\frac{A^4\sigma_{n_r}^2+4\sigma _{m}^{4}\sigma_{n_r}^2-2\sigma _{m}^{2}\sigma_{n_r}^2A^{2}}{8\sigma _{m}^{6}P_t}
\end{aligned}
\end{equation}
As conditional BER of coherent modulation scheme is $P_e(\tau)=\kappa Q(\sqrt{\overline{\gamma}\zeta\tau})$, we can obtain the asymptotic average BER as
\begin{equation} \label{3-7}
\begin{aligned}
P_e^{\infty }&=\int_{0}^{\infty}\kappa Q(\sqrt{\overline{\gamma}\zeta\gamma})f_{\gamma}(\gamma)d\gamma\\
&\approx \frac{\kappa\Gamma{\left(\frac{\rho+1}{2}\right)}}{4\sqrt{\pi}\zeta^{\frac{\rho }{2}}}\left ( \frac{\sigma_{n_k}}{A_0P_t\delta\alpha_m\mu_kh_{l_{r,k}}} \right )^{\rho }e^{2\rho \sigma _{X}^{2}+2\rho ^{2}\sigma _{X}^{2}}\\&+\frac{A^4\sigma_{n_r}^2\kappa+4\sigma _{m}^{4}\sigma_{n_r}^2\kappa-2\sigma _{m}^{2}\sigma_{n_r}^2A^{2}\kappa}{16\sigma _{m}^{6}\zeta P_t}
\end{aligned}
\end{equation}
It is clear from \eqref{3-5} and \eqref{3-7}, the performance of the considered system is dominated by the worst link, which depends on the parameters of the two links.
\section{numerical results} \label{num}
In this section, we compare analytical results and simulation results. The simulation in this paper is based on physical modeling. The RF signal reaches the relay through the Rice channel, and the relay decodes and converts it into an optical signal and forwards it to different ORISs with different power distribution coefficients. The incident light beam is reflected to the receiving plane by the ORIS according to the reflection law. We have added independent jitter random variables to the direction vector of the optical beam and the normal vector of the ORIS surface. We simulated $10^8$ independent signals at the transmitting end, and used Monte Carlo method at the receiving end to count the outage probability and BER. The parameters in the systems are presented in Table \ref{table.1}.\par
\begin{figure}[htbp]
\centering
\includegraphics[width=0.49\textwidth]{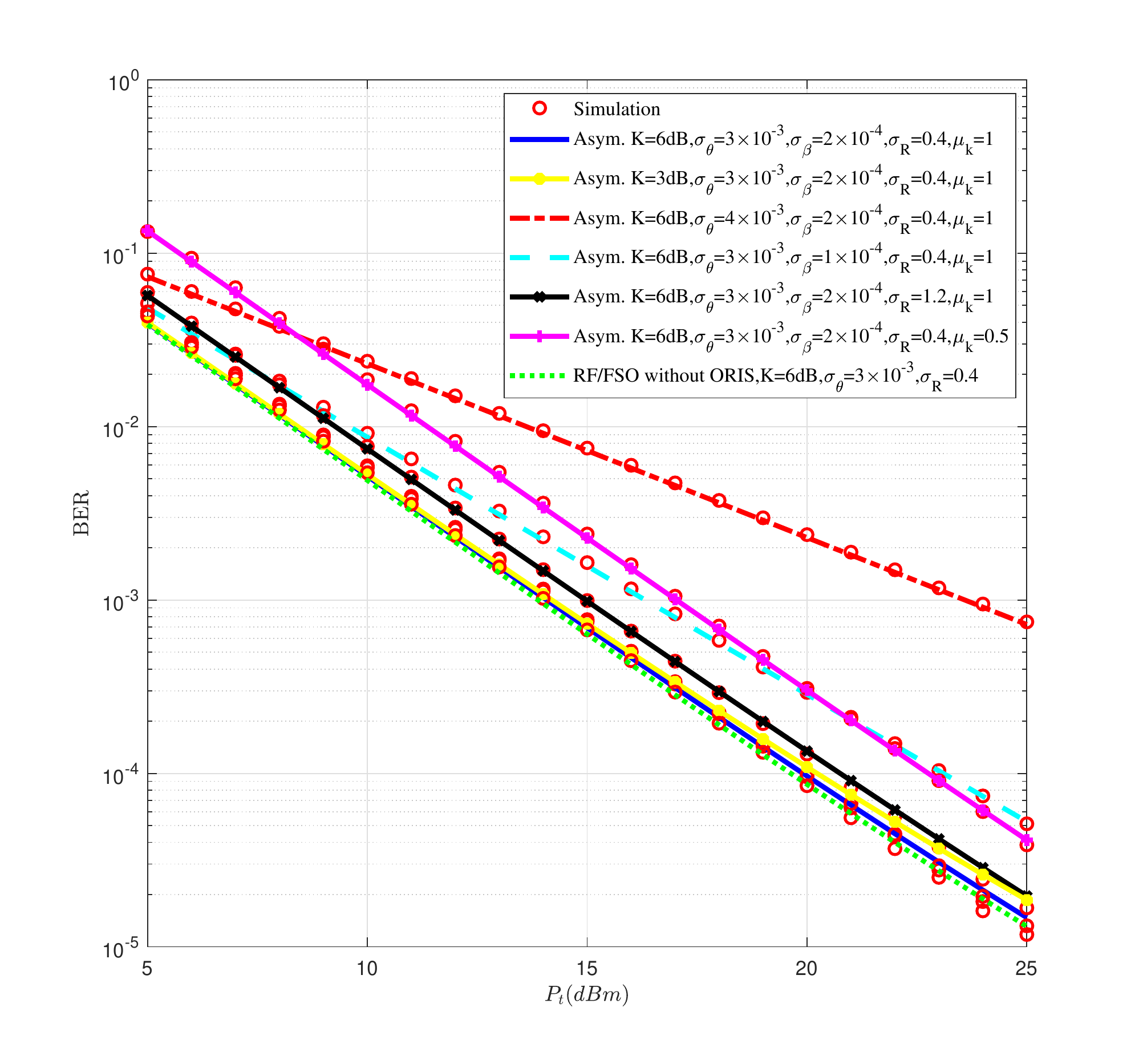}
\caption{The asymptotic BERs and simulated BERs for hybrid RF-ORIS assisted FSO system with different parameter values, the asymptotic results are obtained from \eqref{3-7}.}
\label{fig.2}
\end{figure}
\begin{figure}[htbp]
\centering
\includegraphics[width=0.49\textwidth]{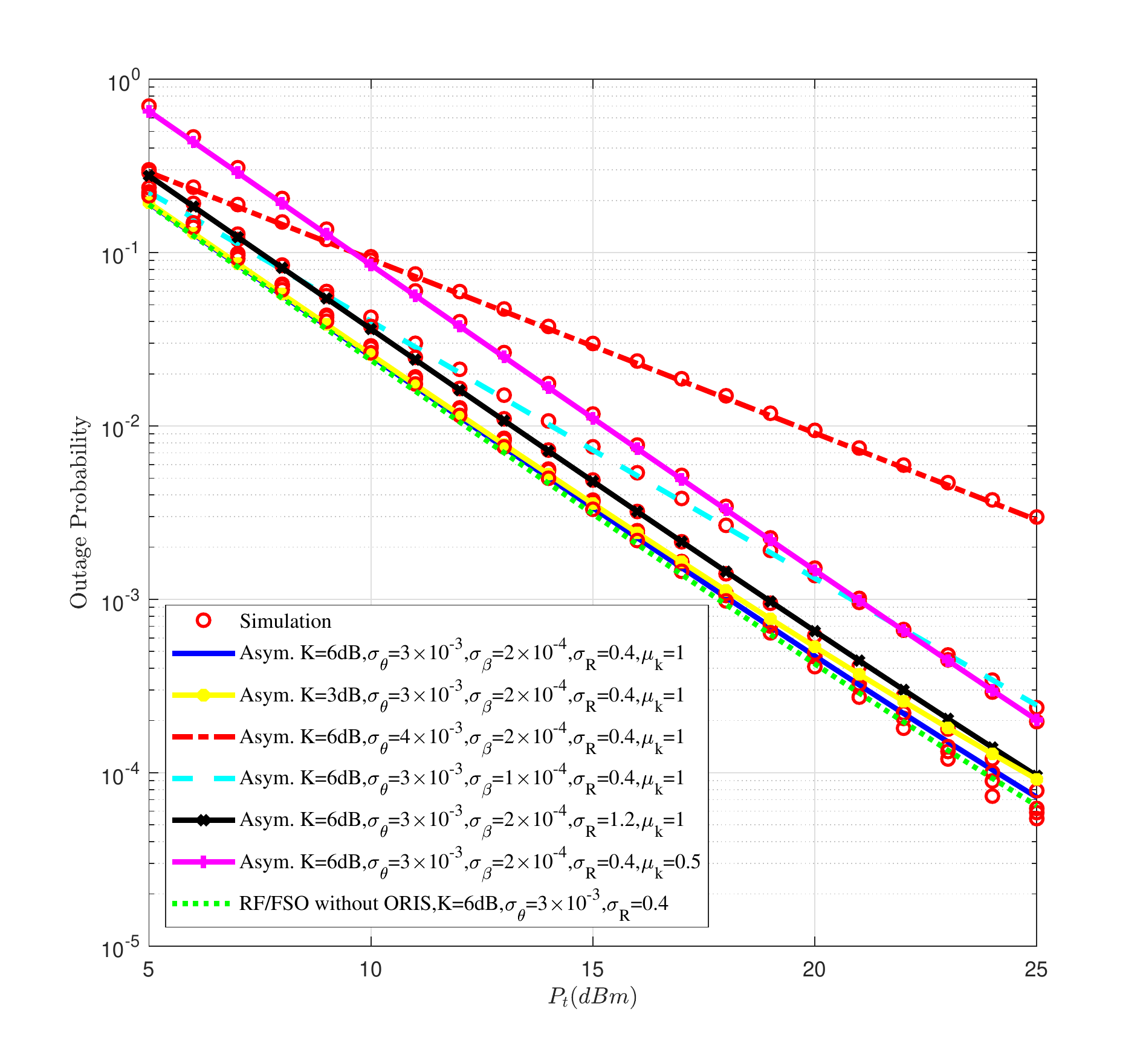}
\caption{The asymptotic outage probability and simulated outage probability for hybrid RF-ORIS assisted FSO system with different parameter values, the asymptotic results are obtained from \eqref{3-7}.}
\label{fig.3}
\end{figure}

In Fig. \ref{fig.2}, we show the asymptotic BERs and the simulated BERs for the hybrid RF-FSO system assisted by ORISs with different parameters. The asymptotic BER curves are based on \eqref{3-7}. The outage probability curves for the same systems with SNR threshold $\gamma_{th}=5dB$ are presented in Fig. \ref{fig.3}, where the asymptotic outage probability curves are obtained by \ref{fig.3}. From Fig. \ref{fig.2}, the simulated BER curves for IM/DD with OOK modulation agree with the asymptotic BER curves in high SNR regimes. The same behavior can be observed for the outage probability in Fig. \ref{fig.3}. The numerical results indicate that the asymptotic estimation of system performance is accurate in large SNR regimes.\par
Comparing with the curve of RF/FSO system without ORISs, we can observe that the RF/FSO links with ORISs show a slightly worse performance than the traditional RF/FSO links. However, the performance degradation introduced by ORIS is relatively small, and at the cost of this, the FSO link obtains a higher degree of freedom of beam adjustment. Comparing the curves with different Rice factor $K$, we can observe that as $K$ decreases, system performance becomes worse, which indicates that the increase in the proportion of deterministic components of the RF signal can improve the performance. However, since this system is a hybrid link, the influence of the Rice factor is relatively small. Comparing the curves with different $\sigma_\theta, \sigma_\beta$ and $\sigma_R$, we can observe that the variety of pointing error will bring about a large impact on performance, while the variety of atmospheric turbulence parameters will have a small impact on system performance. This is because the ORIS's reflection and jitter increases the impact of pointing error on system performance. Meanwhile, the length of the optical link  in this system is only $150 m$, thus the influence of atmospheric turbulence is relatively small. Therefore, in this system, the benefit of correcting or compensating for the fading caused by pointing error is greater than that of atmospheric turbulence. In addition, as $\mu_k=0.5$, which is equivalent to the relay supporting two users in parallel, the BER increases by about $5 dB$ at $P_t=20 dBm$. From the curves, we can observe that $\mu_k$ is not proportional to the outage probability and BER, which indicates that an additional user may introduce a certain performance fading.
\begin{table}[htbp]
\centering
\caption{SYSTEM SETTINGS}\label{}
\begin{tabular}{c|c}
\hline
Parameters& value\\
\hline
Optical wavelength ($\lambda$)& 1550 nm\\
Noise variance at relay ($\sigma_r^2$)& $10^{-4}$ W\\
Noise variance at receiver ($\sigma_{n_k}^2$)& $10^{-4}$ W\\
Optical power conversion coefficient ($\delta$) & $0.8$\\
Transmit divergence at $1/e^2$ ($\phi$)& 8 mrad \\
Corresponding optical beam radius ($w_z$)& $\approx$ 120 cm\\
Optical path loss $h_{l_{r,k}}/l$ & $0.1$ dB/km\\
Link distance from transmitter to relay ($l_{s,r}$)& 100 m\\
Link distance from relay to ORIS ($l_{r,o}$)& 50 m\\
Link distance from ORIS to receiver ($l_{o,u}$)& 100 m\\
Pointing error angle standard deviation ($\sigma_{\theta}$)& 5 mrad \\
ORIS jitter angle standard deviation ($\sigma_{\beta}$)& 2 mrad \\
ORIS attenuation coefficient ($\alpha_m$)& $0.95$\\
Receiver diameter (2a)& 20 cm\\
\hline
\end{tabular}
\label{table.1}
\end{table}

\section{conclusion} \label{conclusion}
In this work, we, for the first time, derive the closed-form solutions of the exact outage probability, asymptotic outage probability and BER of a hybrid RF-ORIS assisted FSO communication system. The RF link is modeled as the Rice fading and the FSO link is analysed based on ORIS model, pointing error and atmospheric turbulence. The hybrid RF-ORIS assisted FSO links are found to show a slightly worse performance as compared to the traditional RF-FSO links. At the cost of this, the FSO link obtains a higher degree of freedom of beam adjustment.
\bibliographystyle{IEEEtran}
\bibliography{IEEEabrv,ref}
\end{document}